\documentstyle[sprocl]{article}

\newcommand{\Avg}[1]{\left\langle #1 \right\rangle}

\def\ol#1{\overline{#1}}

\def\roughlyup#1{\mathrel{\raise.3ex\hbox{$\sim$\kern-.75em
\lower1ex\hbox{$#1$}}}}
\def\roughlydown#1{\mathrel{\raise.3ex\hbox{$#1$\kern-.75em
\lower1ex\hbox{$\sim$}}}}

\def\Avg#1{\langle #1 \rangle}

\def\eqa{\begin{eqnarray}}
\def\eeqa{\end{eqnarray}}
\def\eq{\begin{equation}}
\def\eeq{\end{equation}}

\def\Pl{\gamma_\lft}
\def\lft{{\scrs L}}

\def\sw{s_w}

\def\GF{G_{\scrs F}}

\def\Asl{\hbox{/\kern-.7500em\it A}} 

\def\nubar{\ol{\nu}}

\def\Sca{{\cal A}}

\def\sss{\scriptscriptstyle}

\def\lft{{\sss L}}
\def\sw{s_w}

\def\GF{G_{\sss F}}

\def\dne{\delta n_e}

\begin{document}

\rightline{McGill-99/20}

\title{NEUTRINO PROPAGATION THROUGH FLUCTUATING MEDIA}

\author{C.P. BURGESS}

\address{Physics Department, McGill University, 3600 University Street,\\
Montr\'eal, Qu\'ebec, Canada, H3A 2T8.\\E-mail: cliff@physics.mcgill.ca}

\maketitle\abstracts{The description of neutrino propagation
through a fluctuating medium is summarized.
Fluctuation-Dissipation-Theorem arguments
relate microscopic fluctuations to thermodynamic
quantities, allowing these to be very generally studied in
astrophysical environments. Fluctuation-induced
modifications to neutrino oscillations have been studied
in detail within the Sun and within the envelope of
supernovae, and although surprisingly large effects have
been found none appear to be observable in the near
future.}

\section{Introduction}

The same properties which make neutrinos elusive make them
interact uniquely within astrophysical and cosmological
environments. The incredible feebleness of their interactions
permit neutrinos to escape from deep within stars and supernovae,
making them superb probes with which to potentially `observe'
the interior workings of these objects.

On the other hand, neutrino interactions are not negligible
within these environments since the extreme temperatures and
densities which arise there preclude their neglect.
Despite the potential importance of these neutrino-matter
interactions, some of the approximations inherent in
these analyses have only recently been
systematized\cite{sawyer,n1,n1a,n2,annphys,n4,helio,spinfl,EV}.
It is the purpose of this talk to briefly summarize recent
studies of how neutrinos interact with dense, hot environments.

\section{The Nature of the Problem}

A great deal is known about how photons interact with
matter, and much (but not all) of the intuition developed for photons
can be carried over to the study of neutrinos. We therefore
pause here to review the description of photons propagating
through a medium such as a dielectric.

\subsection{Photons}

At a microscopic level, photons couple to dielectric
materials by interacting with the electrons within
its constituent atoms. The underlying electron-photon interaction is
expressed in terms of the electromagnetic current, $j^\mu$,
and the electromagnetic potential, $A_\mu$,
by the interaction lagrangian density:
\begin{equation}
\label{QED}
{\cal L} = e\,  A_\mu \, j^\mu.
\end{equation}

Although the smallness of $e$ ensures interactions with
any particular atom are weak, the interaction probability becomes unity
once sufficiently many atoms are encountered. For everyday
dielectrics sufficiently many atoms are encountered over
a characteristic distance, $\xi$, which is usually much smaller
than the macroscopic distances over which the propagation of light is to be
understood.

What is generically a complicated, strong-coupling
problem considerably simplifies if the medium's fluctuations
are correlated over distances, $\ell$, which are sufficiently
small in comparison with $\xi$. In this case, although
fluctuations can strongly influence the evolution of electromagnetic
waves, they do not build up correlations over significant distance
scales.

The condition $\ell \ll \xi$ permits simplification
for two reasons.
First, the absence of correlations over distances larger than $\ell$
permits photon evolution to be described by a {\it Markov} process.
That is, it ensures that the evolution may be described by a
differential equation which is local in time and space.
Second, the smallness of $\ell$ compared to $\xi$
permits the resulting differential equation to be evaluated
perturbatively in the microscopic couplings.

When these conditions are satisfied, the interactions of photons
with the medium are well described by
Maxwell's equations, supplemented by a dielectric function
which describes the influence of the medium. Furthermore,
the dielectric function itself may be
computed perturbatively in powers of $e$, with the
fluctuations contributing to the electromagnetic
polarizability:\cite{LL}
\begin{equation}
\label{QEDcorr}
\Pi^{\mu\nu}(x) = e^2 \, \langle j^\mu(x) j^\nu(0) \rangle .
\end{equation}
Notice that the leading term arises at second order in $e$ because, for
dielectrics, the lower-order contribution vanishes
because $\langle j^\mu (x)\rangle = 0$.

A virtue of this kind of formulation is that it separates
issues. The correlation function, eq.~(\ref{QEDcorr}),
is seen to control photon propagation regardless of the
microscopic details of the fluctuations themselves.
The nature of a particular fluctuation only enters once
$\Pi^{\mu\nu}$ is evaluated in order to obtain the dielectric
function explicitly for the system of interest.

\subsection{Neutrinos}

Similar considerations apply to neutrino interactions with
fluctuating media. The microscopic interactions can be
taken to be
\begin{equation}
\label{Fermi}
{\cal L}_{\scriptscriptstyle F} =   i {\GF  \over \sqrt2}\, \sum_k
\left( \nubar_k \gamma_\mu  \Pl \nu_k \right)\;  J^\mu_k(x),
\end{equation}
where the sum on $k$ runs over the three neutrino species and
$J^\mu_k$ describes the contribution of other particles to
the weak interactions.

As before, a perturbative calculations are
possible for fluctuations having sufficiently
small correlation lengths. An important difference in the
neutrino case is that the leading influence of the medium is
described by neutrino coupling to the mean, $\langle J^\mu_k
\rangle$. For
many astrophysical applications the medium of interest is
a parity invariant mix of protons, electrons and neutrons,
for which
\begin{equation}
\label{meanJ}
\langle J^\mu_k(x) \rangle = 2 \delta_{ke} \, j^\mu_e(x) -
j^\mu_n(x) +  \, (1 - 4 \sw^2) \; [j^\mu_p(x) - j^\mu_e(x) ] ,
\end{equation}
where $j_f^\mu$ (for $f = e,p$ and $n$) are the currents which
follow the flow of electron, proton and neutron densities.
This neutrino interaction lagrangian
describes the usual resonant oscillations in matter.

Corrections to this approximation describe neutrino scattering
from the fluctuations about the mean. These are proportional
to the correlation function $\langle J^\mu(x) \, J^\nu(0) \rangle$,
whose size must be evaluated for various fluctuations.

\section{Types of Fluctuations}

Two major kinds of fluctuations are usually examined
when considering particle propagation through a medium:
($i$) equilibrium fluctuations and ($ii$) position-dependent fluctuations,
each of which are now considered in turn.

\subsection{Spatial or Temporal Fluctuations}

The influence of spatial variations in $\langle J^\mu_k(x) \rangle$
(or in the solar magnetic field, $B(x)$) on neutrino oscillations
has recently received considerable study in the literature. Most of these
model the variations as delta-correlated gaussian noise in either
the density of the sun\cite{n1,n2,annphys}, or a supernova envelope\cite{n1a}
or in solar magnetic fields\cite{n1,annphys,EV}.
Solar neutrino evolution through more realistic fluctuations
with long correlation lengths, such as helioseismic waves, has
also been investigated\cite{annphys,helio}.

Although the neutrino-oscillation effects can be surprisingly large,
none of detectable size has yet been discovered.
The difficulty in producing detectable effects can be seen if
the following expression is used, which generalizes to the presence
of fluctuations the Parke formula for resonant neutrino survival
probability:\cite{annphys}
\begin{equation}
\label{parke}
P_e(t,t') = {1\over 2} + \left( {1\over 2} - P_{\sss J} \right) \lambda
\cos 2\theta_m(t')\cos 2\theta_m(t) .
\end{equation}
Here $P_{\sss J}$ is the usual `jump' probability and $\theta_m(t)$
is the matter mixing angle evaluated at the
position occupied by the neutrino at time $t$.

The contributions of fluctuations are summarized
by the coefficient $\lambda$, which is given by
\begin{equation}
\label{lamb}
\lambda\equiv \exp\left[ -2 \, \GF^2 \int_{t'}^t  d\tau \;
\Sca(\tau) \sin^22 \theta_m(\tau) \right] .
\end{equation}

Here $\Sca(t) \equiv\int_{t'}^t  d\tau \; \Avg{\dne(t)\dne(\tau)}$
is the relevant measure of the strength of fluctuations.
An important consequence of this equation
is its implication that the probability of $\nu_e$
survival depends almost exclusively on fluctuation
properties at the position of the MSW
resonance, since $\sin^22\theta_m$ is typically sharply
peaked at this point. It is the difficulty of obtaining large
fluctuations deep within the radiation zone, where the
neutrino resonance takes place, which decisively
suppresses the influence of fluctuations in solar neutrino experiments.

\subsection{Equilibrium Fluctuations}

Equilibrium fluctuations are those within the statistical
ensemble whose average gives the local thermodynamic
properties. For small correlation lengths, and for
nonrelativistic systems, the correlation function of interest
boils down to a measure of local density fluctuations within
the medium.

Such fluctuations are quite generally related by the fluctuation-dissipation
theorem to the equilibrium thermodynamic susceptibilities of  the
system. For instance, in a grand-canonical ensemble local density
fluctuations are related to the system's compressibility:\cite{annphys}
\begin{equation}
\label{fldissthm}
\langle [n(x) - \overline{n}(x)][n(0) - \overline{n}(0)]
\rangle = \left( {\partial n \over \partial p} \right)_T
\, n \, k_{\sss B} T \; \delta^3(x).
\end{equation}

Notice that the generality of this result implies it
works equally well for media which are composed of
weakly- or strongly-interacting systems. So long as they
are in equilibrium, these systems
can differ only through their equations of state. For a
dilute, weakly-coupled system, the ideal-gas law ($p =
n k_{\sss B} T$ applies, and the right-hand-side of
eq.~(\ref{fldissthm}) reduces to $n \; \delta^3(x)$. The
resulting neutrino scattering cross section gives the usual
result proportional to $n$.

More complicated equations of states can change the scattering
rate appreciably. A study of the implications of this expression
for neutrino propagation within supernova cores is
presently underway.

\section*{Acknowledgments}
My thanks to the organizers for the very pleasant environment
and the opportunity to speak. My collaborators on this topic
are Yashar Aghababhie, Peter Bamert,
Steven Horwat and Denis Michaud, and our research funds were provided
by {\sl NSERC} of Canada,
with supplements also provided by {\sl FCAR} du Qu\'ebec
and the McGill Faculty of Graduate Studies.


\def\anp#1#2#3{{\it Ann.\ Phys. (NY)} {\bf #1} (19#2) #3}
\def\apj#1#2#3{{\it Ap.\ J.} {\bf #1}, (19#2) #3}
\def\arnps#1#2#3{{\it Ann.\ Rev.\ Nucl.\ Part.\ Sci.} {\bf #1}, (19#2) #3}
\def\cmp#1#2#3{{\it Comm.\ Math.\ Phys.} {\bf #1} (19#2) #3}
\def\ejp#1#2#3{{\it Eur.\ J.\ Phys.} {\bf #1} (19#2) #3}
\def\ijmp#1#2#3{{\it Int.\ J.\ Mod.\ Phys.} {\bf A#1} (19#2) #3}
\def\jetp#1#2#3{{\it JETP Lett.} {\bf #1} (19#2) #3}
\def\jetpl#1#2#3#4#5#6{{\it Pis'ma Zh.\ Eksp.\ Teor.\ Fiz.} {\bf #1} (19#2) #3
[{\it JETP Lett.} {\bf #4} (19#5) #6]}
\def\jpa#1#2#3{{\it J.\ Phys.} {\bf A#1} (19#2) #3}
\def\jpb#1#2#3{{\it J.\ Phys.} {\bf B#1} (19#2) #3}
\def\mpla#1#2#3{{\it Mod.\ Phys.\ Lett.} {\bf A#1}, (19#2) #3}
\def\nci#1#2#3{{\it Nuovo Cimento} {\bf #1} (19#2) #3}
\def\npb#1#2#3{{\it Nucl.\ Phys.} {\bf B#1} (19#2) #3}
\def\plb#1#2#3{{\it Phys.\ Lett.} {\bf #1B} (19#2) #3}
\def\pla#1#2#3{{\it Phys.\ Lett.} {\bf #1A} (19#2) #3}
\def\pra#1#2#3{{\it Phys.\ Rev.} {\bf A#1} (19#2) #3}
\def\prb#1#2#3{{\it Phys.\ Rev.} {\bf B#1} (19#2) #3}
\def\prc#1#2#3{{\it Phys.\ Rev.} {\bf C#1} (19#2) #3}
\def\prd#1#2#3{{\it Phys.\ Rev.} {\bf D#1} (19#2) #3}
\def\pr#1#2#3{{\it Phys.\ Rev.} {\bf #1} (19#2) #3}
\def\prep#1#2#3{{\it Phys.\ Rep.} {\bf #1} (19#2) #3}
\def\prl#1#2#3{{\it Phys.\ Rev.\ Lett.} {\bf #1} (19#2) #3}
\def\prs#1#2#3{{\it Proc.\ Roy.\ Soc.} {\bf #1} (19#2) #3}
\def\rmp#1#2#3{{\it Rev.\ Mod.\ Phys.} {\bf #1} (19#2) #3}
\def\sjnp#1#2#3#4#5#6{{\it Yad.\ Fiz.} {\bf #1} (19#2) #3
[{\it Sov.\ J.\ Nucl.\ Phys.} {\bf #4} (19#5) #6]}
\def\zpc#1#2#3{{\it Zeit.\ Phys.} {\bf C#1} (19#2) #3}

\end{document}